\documentclass[preprint2,a4paper]{aastex}
\input{crab.defs}
\usepackage{epsf}
\usepackage{emulateapj5}
\usepackage{onecolfloat5}
\shortauthors{Bhat et al.}
\shorttitle{Detection of Crab Giant Pulses}
\def\edpaper{{\rm Bowman et al. (2006)}}
\def\edweb{{\rm http://haystack.mit.edu/ast/arrays/mwa}}

\def\SM{{\rm SM}}

\newcommand{\Wtau}{\mbox{$W_{\tau}$}}

\newcommand{\be}{\begin{eqnarray}}
\newcommand{\ee}{\end{eqnarray}}

\begin{document}
\twocolumn[
\title{Detection of Crab Giant Pulses Using the Mileura Widefield Array Low Frequency Demonstrator
Field Prototype System}
\medskip
\author{N. D. Ramesh Bhat$^1$, Randall B. Wayth$^{2,5}$, Haydon S. Knight$^1$, Judd D. Bowman$^3$, Divya Oberoi$^4$, David G. Barnes$^{1,5}$, Frank H. Briggs$^{6,7}$, Roger J. Cappallo$^4$, David Herne$^8$, Jonathon Kocz$^6$, Colin J. Lonsdale$^4$,  Mervyn J. Lynch$^8$, Bruce Stansby$^8$, Jamie Stevens$^{5,9}$, Glen Torr$^{10}$, Rachel L. Webster$^5$, J. Stuart B. Wyithe$^5$}
\affil{$^1$Centre for Astrophysics \& Supercomputing, Swinburne University, Hawthorn, Victoria 3122, Australia \\
$^2$Harvard-Smithsonian Center for Astrophysics, 60 Garden Street, Cambridge, MA 02138-1516, USA\\
$^3$MIT Kavli Institute for Astrophysics \& Space Research, 77 Massachusetts Ave,
Cambridge, MA, USA\\
$^4$Massachusetts Institute of Technology, Haystack Observatory, Route 40, Westford, MA 01886, USA\\
$^5$School of Physics, University of Melbourne, VIC, 3010, Australia\\
$^6$Australian National University, Research School of Astronomy \& Astrophysics, Weston Creek, Australia\\
$^7$Australia Telescope National Facility, CSIRO, PO Box 76, Epping, NSW 1710, Australia\\
$^8$Department of Applied Physics, Curtin University of Technology, GPO Box U1987, Perth, WA, Australia\\
$^9$Mathematics and Physics, University of Tasmania, Hobart, Tasmania 7005, Australia\\
$^{10}$The Australian National University, Faculty of Science, Canberra, ACT 0200, Australia}
\begin{abstract}
We report on the detection of giant pulses from the Crab Nebula
pulsar at a frequency of 200 MHz using the field deployment
system designed for the Mileura Widefield Array's Low Frequency
Demonstrator (MWA-LFD). Our observations are among the first
high-quality detections at such low frequencies.
The measured pulse shapes
are deconvolved for interstellar pulse broadening, yielding 
a pulse-broadening time of 670$\pm$100 $\mu$s, and the implied 
strength of scattering (scattering measure) is the 
lowest that is estimated towards the Crab nebula from 
observations made so far. The sensitivity of
the system is largely dictated by the sky background, and our
simple equipment is capable of detecting pulses that are brighter than
$\sim$9 kJy in amplitude. The brightest giant pulse detected in
our data has a peak amplitude of $\sim$50 kJy, and the implied
brightness temperature is $10^{31.6}$ K. We discuss the giant
pulse detection prospects with the full MWA-LFD system. With a
sensitivity over two orders of magnitude larger than the
prototype equipment, the full system will be capable of 
detecting such bright
giant pulses out to a wide range of Galactic distances; 
from $\sim$8 to $\sim$30 kpc depending on the frequency. 
The MWA-LFD will thus be a highly promising instrument for 
the studies of giant pulses and other
fast radio transients at low frequencies.
\end{abstract}
]


\section{Introduction} \label{s:intro}

Giant pulses from the Crab Nebula pulsar have been known since the 
early days of pulsar observations 
(Staelin \& Reifenstein 1968; Argyle \& Gower 1972).
These sporadic, large-amplitude bursts can exceed mean pulse energies
by factors as large as $10^4$ \citep{lundgrenetal1995}, and they remain 
as an intriguing phenomenon for theories of the pulsar emission 
mechanism to explain. Their short durations also allow them to serve as 
sensitive probes of the nebula and the interstellar medium. Much has been 
learnt from observations so far, and it is fairly well established that: 
(i) the fluctuations in their amplitudes are associated with changes in 
the coherence of the radio emission \citep{lundgrenetal1995}, (ii) the 
emission is broadband -- extending over several hundreds of MHz 
\citep{sallmenetal1999,popovetal2006}, and (iii) they are superpositions 
of extremely narrow nanosecond-duration structures \citep{hankinsetal2003}. 
The Crab giant pulses have some of the largest implied brightness 
temperatures of known astrophysical sources, and are known to 
reach $10^{32}$ K at microsecond time resolutions \citep{cordesetal2004}, 
and even higher ($10^{37}$ K) at nanosecond resolutions \citep{hankinsetal2003}.

Although the Crab pulsar was originally discovered at a low radio 
frequency of 111 MHz \citep{staelin1968} through detection of its 
giant pulses, the bulk of the observational studies of it over the past 
three decades has focused primarily at higher frequencies
\citep[e.g.][]{moffett-hankins1996,cordesetal2004}. 
This is partly due to limitations imposed by the dispersive and
scattering effects in the interstellar medium (ISM) that
degrade the signal strength at low frequencies. 
A more significant reason perhaps is that most existing large
instruments are optimised for best performance above 300 MHz. For
single dishes this has been driven by a quest for sensitivity, and for
interferometers by resolution and sensitivity, and also the need to 
avoid the problems due to ionospheric distortion of astronomical 
signal at low frequencies. The recent enormous increase in affordability of
computing and the impressive advances in the capabilities of digital
hardware have brought the computational needs of low frequency radio
imaging within reach. This has led to a resurgence
in the interest in low frequency astronomy and a number of instruments
like the Mileura Widefield Array Low Frequency Demonstrator
(MWA-LFD) in Western Australia, Low Frequency Array (LOFAR) in the 
Netherlands and Long Wavelength Array (LWA) in South-west US are 
currently in varying stages of design, development and deployment. 
The ever growing problem of man made radio frequency interference (RFI) 
continues to remain a major concern however, especially in densely populated 
parts of the world.

Despite the paucity of low-frequency studies of the Crab, a wealth of 
information can be gleaned from such studies. As such pulsar radio 
emission is strongly frequency dependent, with most intrinsic and 
extrinsic effects becoming more pronounced at
lower radio frequencies. In particular, the Crab pulsar is well
known for its highly peculiar and complex frequency evolution 
of its pulse morphology and structure
\citep{cordesetal2004,moffett-hankins1996}. Therefore it would be
interesting to examine how the properties of pulse emission
evolve at low radio frequencies. Further, the scattering and
dispersion effects also become more prominent with decreasing 
observing frequency, and are best studied at low frequencies.

The MWA-LFD is a promising instrument with capabilities
that will complement those of existing large telescopes,
particularly the frequency coverage. Given its high sensitivity, 
multi-beaming and wide field-of-view capabilities, this 
instrument promises excellent prospects for pulsar and 
transient science. 
In this paper, we report on successful detection of the 
Crab giant pulses at 200 MHz using the prototype system
that was built for its early deployment activities. 
Although the Crab pulsar was detected earlier at a 
nearby frequency (196.5 MHz) via its average emission 
\citep{rankinetal1970}, our observations mark the first 
detection of individual giant pulses at 200 MHz,
and the data are used to estimate the scattering due to 
the intervening ISM as well as the pulse amplitudes and 
brightnesses. We also assess the detectability of giant 
pulses with the full MWA-LFD system.

\section{Observing System and Data Processing}

\subsection{The MWA-LFD Early Deployment system}

The MWA-LFD array design covers a broad frequency range from 80 to
300 MHz, and features a large number of small phased-array antenna 
systems or {\it tiles}, each consisting of a 4 x 4 array 
of dual-polarisation wideband active dipoles and an analog
delay-line beam-former unit. These tiles yield a field of view of 
$\sim$200$\lambda^2$ ${\rm deg^2}$ (where $\lambda$ is the 
observing wavelength in meters), i.e., approximately $20^{\circ}$ in
angular extent at a frequency of 200 MHz. Three such tiles were 
deployed at the Mileura station in Western Australia 
($26^{\circ}26'S, 117^{\circ}12'E$) as part of the ``early
deployment'' (hereafter ED) program, with baselines of 
$\approx$146 m, $\approx$325 m and
$\approx$212 m. These tiles, plus a simple data capture and 
software correlation system, form 
the observing system used for our data collection. Further details 
on the ED system and activities are
described in \edpaper. For an overview of the array design and 
science goals see \edweb. The
full system, when operational, will consist of 512 such antenna 
tiles, more than 95\% of which will 
be distributed in an area $\sim$1.5 km in diameter. The remaining
tiles will be distributed in a concentric ring with an outer 
diameter of $\sim$3.0 km. The entire band from 80 to 300 MHz 
will be sampled, 32 MHz of which will be processed and recorded.

\subsection{Observations of the Crab Pulsar}

Observations of the Crab pulsar were made on 22 September 2005,
during the last of four two-week long expeditions conducted as 
part of the ED program.  Data were recorded over a bandwidth of 
8 MHz (approximately 6 MHz of useful bandpass)
centred at frequencies 100 and 200 MHz for a total duration of
approximately 3.5 hours each.
The signal was digitally sampled at 16 MHz using the ``Stromlo
Streamer'' \citep{briggsetal2006}
using 8 bits and then re-encoded into 4 bits using a
scheme similar to ``$\mu$-law'' companding. This scheme uses
increasingly
larger spaces between quantisation levels which allows the
input power to
vary by several orders of magnitude without incurring
significant losses in
sensitivity.

Four independent data streams were recorded at a given time.
The system was configured to record one of the two orthogonal 
polarizations from two tiles, and both polarizations from the 
third tile. This implied an aggregate data rate of 32 mega 
bytes per second, i.e., a raw data volume of $\approx$0.8 
tera bytes. The data were stored on disks for offline 
processing and all processing operations, such as detection 
and de-dispersion, as well as generation of spectrometer 
(filterbank) data and phased-array data streams,
were performed in software using the Swinburne supercomputing 
facility.

\subsection{Data Processing and Analysis}

Data processing was performed in multiple stages. The initial
processing was done using software that was developed for 
calibration of the ED system. This was tailored to output an 
incoherent filterbank stream of data at a time resolution of 
1024 $\mu$s and a spectral channel resolution of 8 kHz, 
extracting only the central 4 MHz part of the band.
This coarse resolution processing, which results in significant
dispersive smearing\footnote{The dispersive delay, 
${\rm t_{dm} \approx 8.3 ~ {\mu}s~DM~\Delta \nu/ \nu^3}$
where DM is the dispersion measure, $\Delta \nu$ is the 
spectral channel width (MHz) and $\nu$ is the frequency 
(GHz) of observation.} ($\approx$471 $\mu$s), served as 
the preliminary processing 
pass in order to examine it for visible signatures of giant 
pulses before embarking on a more detailed and rigorous analysis.
Though it limited our sensitivity to only the brighter giant
pulses, it still proved very useful.

The data were then subjected to a blind search for giant pulses
by performing standard procedures such as dedispersion and 
progressive smoothing of the dedispersed time series, and 
identifying samples that are above a set threshold. 
Dedispersion was performed at
multiple trial values around the Crab's
nominal dispersion measure of 56.791 \dmu in order to confirm
the detections. This analysis yielded a handful of confirmed detections at 200 MHz
and none at 100 MHz. The data at 200
MHz were subsequently reprocessed to output a filterbank stream
at a much higher time resolution of 256 $\mu$s and an improved 
spectral resolution of 4 kHz. 
A similar giant pulse search procedure was then repeated. 
The implied
dispersive smearing in this case (235 $\mu$s)
is quite comparable to the time resolution. 
This procedure confirmed all the detections from the
first coarse-resolution search, but
did not result in any additional detections.

In order to overcome the inherent limitations in processing
such incoherent filterbank data, we developed software that 
directly operates on raw voltage samples and performs a 
phased-array addition of the signals, eventually constructing 
a coherent
filterbank stream of data\footnote{Data are coherently 
dedispersed prior to forming a filterbank stream.}.
Details of this scheme are described in the following two 
sections.

\subsubsection{Generation of Phased-array Data}

The underlying principle here is similar to that of the
{\it tied array} mode of operation as employed in 
instruments such as the Australia Telescope
Compact Array (ATCA) and the Westerbork Synthesis Radio 
Telescope (WSRT), or the phased-array mode at
the Giant Metrewave Radio Telescope (GMRT). Here, 
voltage samples from individual array elements
are summed after correcting for
geometric and instrumental delays
in arrival times of the signals. For our ED system, 
the combination of rather short baselines and long wavelengths
means such corrections are largely dominated by the array 
layout or geometry, rather than any
tile-dependent phase corrections as derived from the calibration 
process. In fact, applying the
latter results in only a marginal improvement in the signal 
strength. While instruments such as
the ATCA and the WSRT are equipped with dedicated hardware to 
perform such an operation, in our case, this phased-array 
addition of signals was fully realized in software.

For our 3-tile ED system, we would expect the signal to be
improved by a factor of $\sqrt{3}$ in this manner, but in
practice even more by performing a coherent dedispersion of
the signal. Ideally, the phased array summation will need to be
performed separately on each of the two orthogonal polarizations, 
however in our case only
the east-west polarization has signals recorded for all three 
tiles. The phased-array data
stream after the summation is simply written out as floating
point numbers, mainly for the simplicity of the data format 
and compatibility with the processing software. This resulted 
in a factor of four increase in the net data volume ($\approx$1.5 
tera bytes per frequency). The data were transfered onto disk 
drives and made accessible to the Swinburne supercomputer
via the Grange-Net facility.

\begin{figure}[t]
\epsscale{1.0}
\plotone{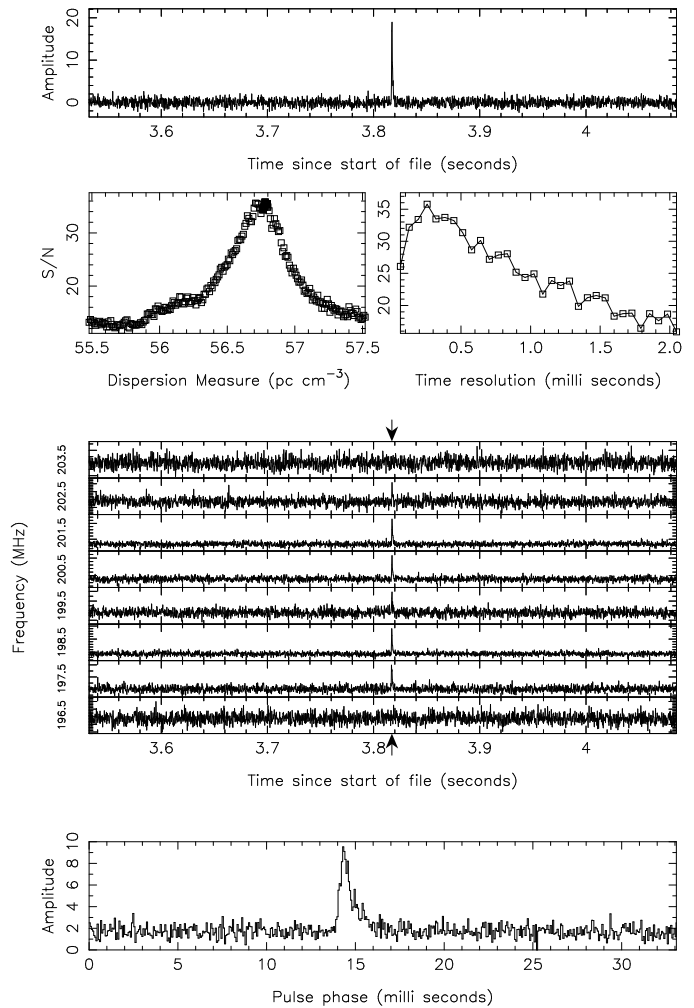}
\caption{A bright giant pulse detected with the MWA-LFD early
deployment system.
{\it Top}: Dedispersed time series of a short data segment near
the giant pulse.
{\it Middle panels}: Diagnostic plots from the giant pulse
search processing; 
plots of signal-to-noise ratio vs dispersion measure and time
resolution, and
dedispersed time series for 8 sub-bands, each 1 MHz wide, within the 8
MHz bandwidth.
{\it Bottom}: The dedispersed pulse shown in high time resolution. The
scattering tail is clearly visible.
}
\label{fig:bestgp}
\end{figure}

\subsubsection{Dedispersion and Giant Pulse Search}

Our phased-array data streams were written out in the form of
voltage samples (i.e., baseband data format), and hence allowed 
a coherent dedispersion to be
performed on the signal.
This procedure removes the dispersion entirely, and is 
especially important at low
radio frequencies where the signal degradation due to 
dispersion can be rather severe. The
dedispersed data are then subjected to
the pulse detection algorithm to search for giant pulses. The
procedures we have adopted are very similar to that described 
in \citet{knightetal2006}. The
voltage samples are first Fourier transformed to the frequency 
domain and the spectra are divided
into a series of sub-bands. Each sub-band is then multiplied 
by an inverse response filter
(kernel) for the ISM \citep[e.g.][]{hankins-rickett1975}, and 
then Fourier transformed back to the time domain to construct 
a time series with a time resolution coarser than the
original data. By splitting the input signal into several 
sub-bands, the dispersive smearing
is essentially reduced to that of an individual sub-band. This
also means
the procedure uses shorter transforms than single-channel
coherent dedispersion
and is consequently a more computationally efficient method.

This ``coherent filterbank'' stream of data are then square-law
detected and combined after correcting for dispersive delays 
between the sub-bands, to construct a single
coherently dedispersed time series for the entire band. 
The pulse detection procedure
involved progressive smoothing (convolution) of time series 
with matched filters of widths
ranging from 64 $\mu$s to 2048 $\mu$s in steps of a factor 
two, and identifying the intensity samples
that exceed a set threshold (e.g. 8.5 $\sigma$). In addition 
to performing dedispersion at
the Crab's nominal DM of 56.791 \dmu, we also perform this 
procedure over a large
number of adjacent DM values (typically over a DM range 
$\approx$0.1${\rm DM_{crab}}$, in steps of 0.01 \dmu), in 
order to confirm the dispersion signature. For each DM and 
the matched filter width,
we compute signal-to-noise ratio (S/N) of the pulse amplitude 
over a short stretch of time centered on the pulse.
This process is computationally intensive and was carried 
out on Swinburne's supercomputer. From this analysis, 
diagnostic plots are generated for
each candidate giant pulse as shown in Fig.~\ref{fig:bestgp}.
We also check if the signal is broadband by displaying the
dedispersed time series in 8 x 1 MHz sub-bands.
These diagnostic plots were subjected to a careful 
human scrutiny to discriminate real giant pulses from spurious
signals.

\subsubsection{System Sensitivity and Flux Calibration}
\label{s:calib}

The Crab Nebula is a fairly bright and extended source in the
radio sky, with a flux density of $\sim$955$~\nu^{-0.27}$ Jy 
\citep{bietenholzetal1997} and a characteristic diameter of 
$\sim$5$^{\prime}$.5. Thus, in general, for most conventional
observations with large single-dish telescopes such as Arecibo 
and Parkes, the nebular emission
dominates the system noise temperature
\citep[e.g.][]{cordesetal2004}. However, in our case the
combination of a low antenna gain, large sky background, 
high system temperature, large field of view, and 
significant side lobes of the antenna response function,
leads to a very different scenario whereby the system 
temperature \Tsys is heavily dominated by the sky 
background itself. This is further compounded by the 
fact that the tile response
pattern becomes more complex at large zenith angles that 
are relevant for the Crab, and varies
significantly over the duration of the observation 
depending on the zenith angle (ZA). Further
technical aspects of the ED system are discussed by \edpaper.

In order to account for these effects, and to obtain a
realistic estimate of the sky background temperature, 
\Tsky, we convolve the sky model, \Tsky($\theta$),
with the antenna power response function, W($\theta$). 
The sky model \Tsky($\theta$) is obtained by extrapolation of
the 408 MHz sky map of \citet{haslametal1982} assuming 
a spectral index $\alpha$ = $-2.6$.
For our purpose, \Wtheta is the phased-array beam pattern when
the tile beam is pointed toward the Crab.
Performing the above calculation yields \Tsky = 180 K when the 
Crab is at transit (ZA = 48$^{\circ}$), but may increase as 
much as by 20\% at larger zenith angles of the Crab (e.g. 
ZA = 60$^{\circ}$). The net system temperature is given by 
\Tsys = \Tsky + \Trecv,
where \Trecv is the receiver temperature. For the ED system, 
the receiver temperature is a strong
function of frequency, and is $ 200\pm20 $ K near 200 MHz 
\citep{bowmanetal2006}. Thus \Tsys = 380$\pm$30 K at
200 MHz, taking into account the nominal uncertainty in 
receiver temperature measurements
and change in \Tsky with the ZA.

In order to translate our system temperature estimates to
their equivalent flux densities (i.e., the system 
equivalent flux density, \Ssys), we require an estimate 
for the array
gain, G, expressed in conventional units of \kpjy. This 
gain parameter can be derived from the effective area of
the tiles, \Aeff, using the simple relation, 
$ {\rm G = \Aeff / 2 ~ \kB } $, where \kB is the Boltzmann constant. For
the MWA-LFD antenna, the effective area is a strong function of 
frequency and is given by $ \Aeff =
\lambda^2 / x $, where $x$ is the frequency-dependent scaling 
parameter. Based on our design simulations $x=2.1$ at
200 MHz, i.e., $ \Aeff \approx 1~{\rm m^2}$ for a single dipole. 
Thus for the ED system which comprises 48 dipoles
in total, the net effective area is $\approx$ 50 ${\rm m^2}$ at
200 MHz, i.e., an equivalent system gain of $\approx$0.02 \kpjy. 
However, as described in \edpaper, the tile gain also has a
strong functional dependence with the zenith angle. 
The measured dependence of ${\rm cos^{2.6}(ZA)}$ is quite
steeper than a theoretically expected ${\rm cos^{2}(ZA)}$
form, and thus the nominal gain for the ED system is
$\approx$0.008 \kpjy towards the Crab. 
Further, the zenith angle varies from $60^{\circ}$ (near rise
or set) to $45^{\circ}$ (at transit) for the Crab, leading to 
a gain variation by a factor 2.5 over the observation.
Thus, our \Tsys estimate of 380 K corresponds
to an equivalent system noise of \Ssys = \Tsys/G $\approx$ 47.5 kJy.
For our recording bandwidth of 8 MHz and assuming a typical 
matched filter width of 250 $\mu$s, the
effective \Ssys is of the order of $\sim$1100 Jy, i.e.,
a minimum detectable pulse amplitude of $\sim$5.5 kJy for a
5-$\sigma$ detection threshold.

\begin{figure}[t]
\epsscale{1.0}
\plotone{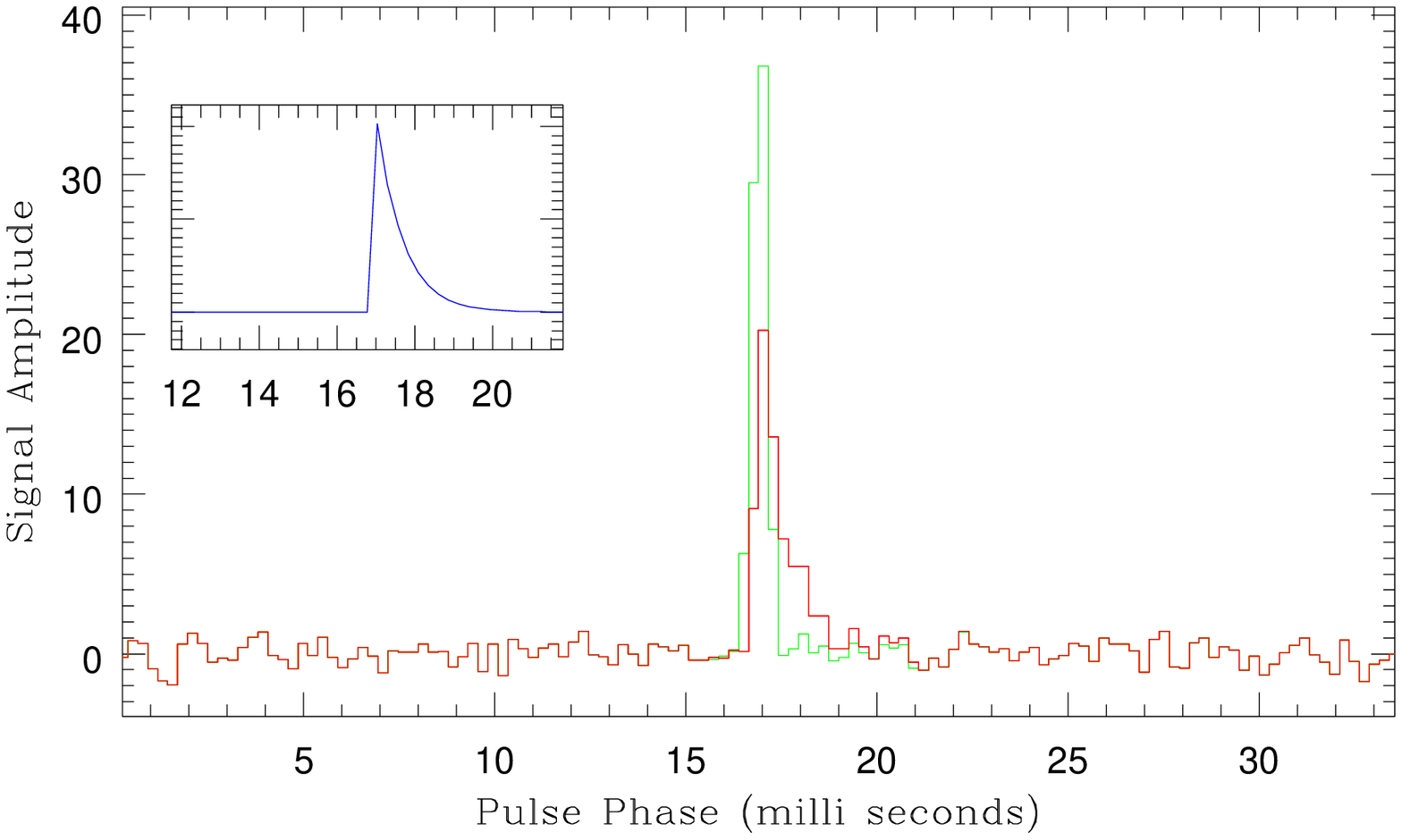}
\plotone{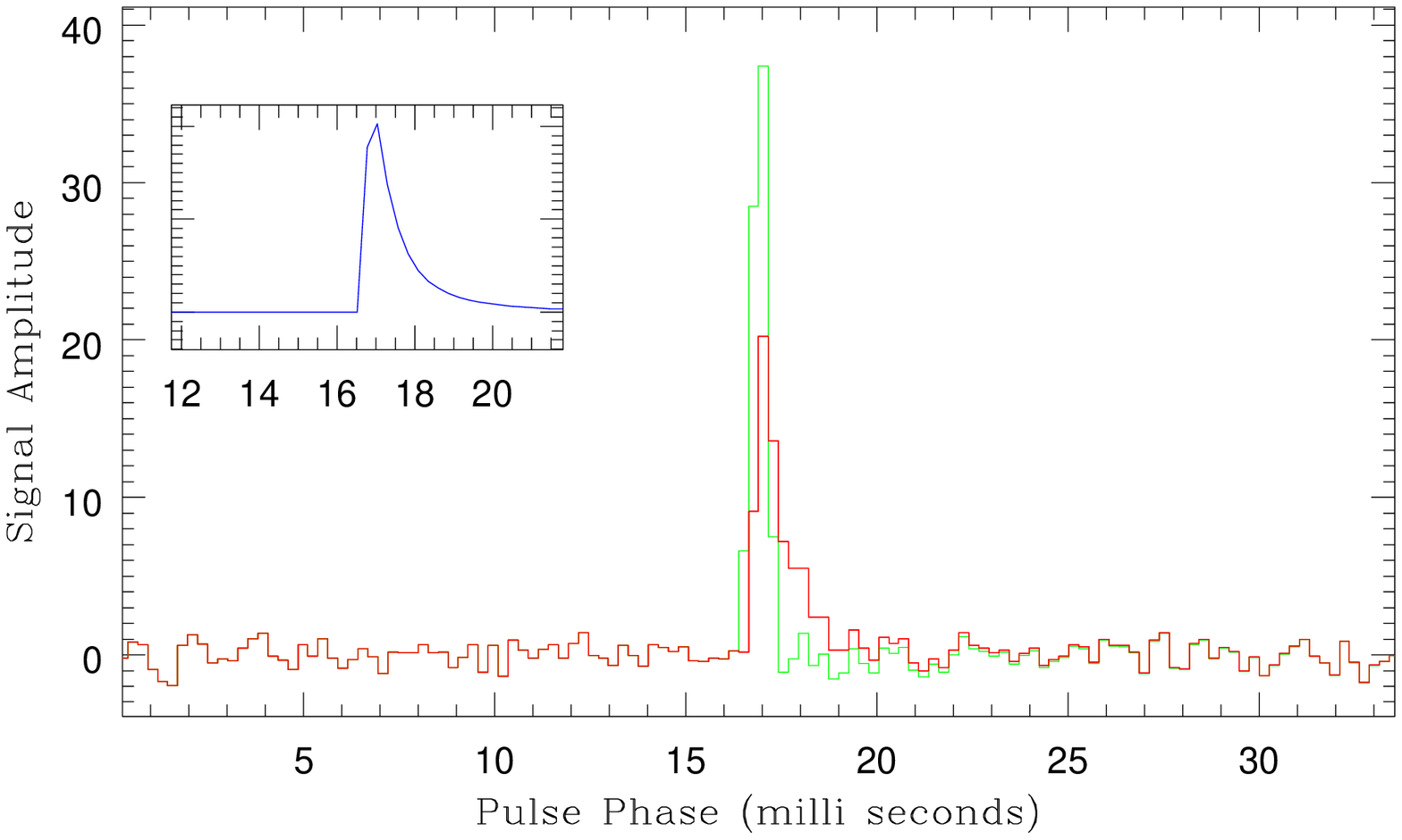}
\caption{Measured and reconstructed pulse shapes of a bright giant pulse at 200 MHz. The best fit 
scattering function (i.e., the pulse broadening function, PBF) is shown in the inset. The top and 
bottom panels correspond to different models of the PBF; an exponential PBF that corresponds to an 
infinite thin screen model of scattering medium ({\it top}), and a more rounded PBF that corresponds 
to a uniform distribution of scattering medium ({\it bottom}). The deconvolution results in a 
significant increase in  the peak signal strength and a pulse shape that is presumably closer to intrinsic.}
\label{fig:clean}
\end{figure}

\subsubsection{Deconvolution of Pulse Broadening}

Fig.~\ref{fig:bestgp} shows an example of giant pulse
detection. As evident from this figure, the measured 
pulse shape is quite asymmetric and is marked by a 
clear lengthening at the trailing end, a characteristic 
of the {\it pulse broadening} phenomenon that results from 
multipath scattering in the intervening interstellar medium (ISM)
\citep[e.g.][]{williamson1972,cordes-lazio2001}. This effect is
more pronounced in observations of distant pulsars and at low
frequencies \citep[e.g.][]{bhatetal2004}. The measured pulse
shape, $y(t)$, can be modelled as effectively a convolution of 
the intrinsic pulse shape, $x(t)$, with the impulse response 
of the scattering medium, $g(t)$, often referred to as the pulse
broadening function (PBF). More precisely,
\be
y(t) = x(t) \otimes g(t) \otimes r(t)
\ee
where $r(t)$ is the net instrumental resolution function.
The observed pulse broadening is usually quantified
by a time scale, \taud, characteristic of a PBF fit to the
measured pulse shape. The exact form of
the PBF and its scaling with frequency depend on the nature of
distribution of scattering material along the line-of-sight and 
on its wavenumber spectrum \citep[e.g.][]{lambert-rickett1999}.

In order to estimate the pulse broadening, we adopt the
CLEAN-based deconvolution approach developed by 
\citet{bhatetal2003}. Unlike the traditional 
frequency-extrapolation approach 
\citep[e.g.][]{lohmeretal2001}, this method makes no 
prior assumption of the intrinsic pulse
shape, and thus offers a more robust means of 
determining the underlying PBF, and therefore
can be applied to a wide variety of pulse shapes and 
degrees of scattering. The procedure involves
deconvolving the pulse shape in a manner quite similar 
to the CLEAN algorithm used in synthesis
imaging, while searching for the best-fit PBF and 
recovering the intrinsic pulse structure. It
relies on a set of figures of merit that are defined 
in terms of positivity and symmetry of the resultant
deconvolved pulse and some parameters characterizing 
the noise statistics in order to determine the
best-fit PBF. Such an approach is especially justified 
for the Crab giant pulses, as they are
known to show structures at timescales down to 
microseconds or even nanoseconds, and 
show complex and quite unusual evolution in pulse 
morphology with frequency.

The deconvolution procedure employed here is similar to that
described in \citet{bhatetal2003}, except that the restoring 
function used in our case is a much
simpler one. This is because the pulse smearing due to 
residual dispersion and instrumental effects is
negligible. The major contributing factor to our restoring 
function is that due to time binning of
the pulse profile. We used two different trial PBFs; the 
first, \gone, is appropriate for a
thin slab scattering screen of infinite transverse extent 
(\PBFone), and the second, \gtwo, corresponds to
a uniformly distributed scattering medium (\PBFtwo). Their 
functional forms are given by
\citep[e.g.][]{williamson1972},
\be
g_1 (t) = \tau _d^{-1}\exp(-t/\taud) \, u(t), \qquad \qquad \qquad
\\
g_2 (t) =( \pi^5 \tau _d ^3 / 4 t^5 )^{1/2} \exp (- \pi^2 \taud / 4t) \, u(t).
\label{eq:pbfs}
\ee
where $u(t)$ is the unit step function, $u(<0) = 0, u(\ge 0) =
1$. While \PBFone, which has a one-sided exponential form, has
been commonly used by several authors in
the past, the latter PBF is a generic proxy for more realistic
distributions of scattering material. The best-fit PBFs 
obtained in this manner along with the
respective intrinsic pulse shapes are shown in 
Fig.~\ref{fig:clean}. As evident from this
figure, the deconvolution process naturally leads to a 
significant increase (nearly by a
factor two) in the effective signal-to-noise ratio of 
the pulse.

\section{Results and Discussion}

\subsection{Giant Pulse Amplitudes and Brightness Temperatures}

The measured amplitudes of giant pulses at low radio
frequencies are determined by a number of factors such as the
nature of the emission spectrum, the degree of scattering and
the time resolution of the data. At frequencies $\ga$ 300 MHz, 
the pulse energies are known to follow a $\nu^{-4.5}$ form, 
i.e., a spectral form that is much steeper than typical 
pulsar spectrum. The behaviour of the spectrum at low 
frequencies is not well understood, although some of the 
early work suggests a turnover of the spectrum near our 
observing frequency of 200 MHz \citep[e.g.][]{manch-taylor77}.
Further, although the Crab pulsar is known to show quite a
peculiar evolution in frequency in terms of pulse morphology
and structure of both the average and giant pulse emission, at
frequencies $\la$ 300 MHz, scattering is likely to dictate the
measured pulse shape as the pulse smearing becomes severe.
Thus, even though the giant pulse emission inherently 
possesses structures down to microsecond or even nanosecond
levels, finer structural details of individual giant pulses are
simply not discernable due to severe broadening of the
intrinsic pulse expected at these frequencies. Although in
principle our deconvolution approach is useful for potentially
extracting the information in such situations, unfortunately
the signal-to-noise of our data does not seem large enough to
attempt such an analysis. The effective time resolution of the
data is also an important factor, as any level of smoothing
(convolution) in time performed to aid either a detection of
pulse, or an enhancement in the signal strength, would mean an
inevitable reduction in the peak pulse amplitudes. Thus, in
short, measurements of giant pulse amplitudes and fluxes at low
frequencies should be treated with some caution, especially 
when compared with similar measurements made at higher frequencies.

\begin{figure}[t]
\epsscale{1.0}
\plotone{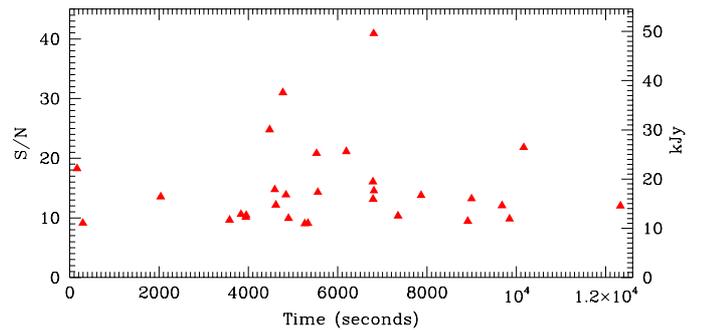}
\caption{Summary plot of giant pulse detections at 200 MHz. The
peak flux densities
range from 10 to 50 kJy. Despite small-number statistics, these
detections seem to
follow a roughly power-law distribution in pulse amplitude and
energy. In total, 31 pulses were detected in 3.5 hours of observations. The
S/Ns shown are for the east-west polarisation; the signal is detected in 
north-south polarisation in most cases, albeit with a poorer S/N.}
\label{fig:giants}
\end{figure}

Results from our giant pulse search are summarised in
Fig.~\ref{fig:giants} where the measured pulse amplitudes 
are plotted against their 
times of occurrence.
The estimates of S/N
are converted to units of Jy using the flux calibration
procedures detailed in \S~\ref{s:calib}. Note that these 
are the measured peak pulse amplitudes, i.e., no
correction is applied for the enhancement in
signal strength achievable by the deconvolution of scattering.
Despite the small-number statistics,
the pulse energies seem to follow a roughly power-law type
distribution (index = $-1.7\pm0.1$). For the nominal
system temperature and gain estimates as described in
\S~\ref{s:calib} (380 K and 0.01 \kpjy respectively), and 
given our typical processing parameters (a net bandwidth of 6
MHz and a nominal pulse width of 300 $\mu$s), the effective 
system noise turns out to be of the
order $\sim$1100 Jy. This means our
system is sensitive to pulses brighter than $\sim$9 kJy,
assuming a 8.5-$\sigma$ detection threshold.
Our search analysis has yielded a total of 31
giant pulses above this
threshold. The brightest giant pulse detected in our data has a
peak flux density of 50 kJy and
and a width $\approx$300 $\mu$s. The measured amplitudes can
be translated to implied brightness
temperatures (\Tb) using a simple relation (based on the
light-travel size and ignoring relativistic dilation),
\be
\Tb = \left( { \Snu \over 2 \, \kB } \right) \left( { {\rm D} \over \nu \delt } \right)^2,
\label{eq:bright}
\ee
where \Snu is the peak flux density and \delt is the pulse
width; \kB is the Boltzmann constant and D corresponds to 
the Earth-pulsar distance. Assuming a peak
pulse amplitude of 100 kJy (i.e., the
amplitude of the reconstructed pulse in Fig.~\ref{fig:clean})
and a nominal distance of 2 kpc to the Crab Nebula 
\citep[e.g.][]{ne2001}, we estimate a 
brightness temperature of $10^{31.6}$ K for our brightest 
giant pulse.
However this is likely to be an underestimate of the true brightness
temperature given the degradation of the signal strength
expected due to pulse broadening. Interestingly, this estimate is 
comparable to that 
derived from the brightest giant pulse  detected at 
430 MHz using Arecibo \citep{cordesetal2004} where
the scattering is at least an order of magnitude 
smaller. Indeed much higher brightness temperatures
($10^{37}$ K) have been reported from observations
at 5.5 GHz (where the scattering is negligibly small) 
made at nanosecond time resolutions
\citep{hankinsetal2003}. Despite the various
factors that degrade the signal strength at low frequencies,
our observations confirm fairly high
brightness temperatures for the giant pulses at 200 MHz.

\begin{figure}[]
\epsscale{1.0}
\plotone{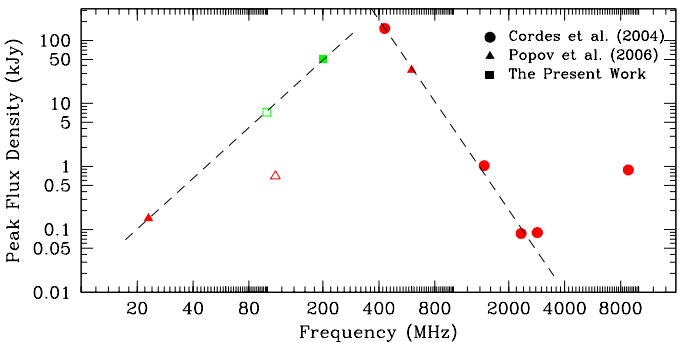}
\caption{Peak flux densities of the brightest giant pulses (in
1 hr) over a wide range
of frequencies from 23 MHz to 8.8 GHz (solid symbols). The
symbols indicate data from different sources -- circles denote 
Arecibo observations of
\citet{cordesetal2004}, triangles are recent low frequency 
observations of \citet{popovetal2006}, and squares
denote the present work. The unfilled symbols near 100 MHz
indicate limits derived from a non-detection with MWA-LFD 
(square) and the brightest giant pulse in 15 minutes of 
observation (triangle).}
\label{fig:brights}
\end{figure}

The amplitude of the brightest giant pulse detected in our data is
compared with similar estimates from
recent observations in Fig.~\ref{fig:brights}. The plot spans a
wide range of frequencies from
$\sim$20 MHz to $\sim$9 GHz, i.e., nearly 3 orders in
magnitude. Observations at most of the higher
frequencies are from an intensive multifrequency campaign with
Arecibo \citep{cordesetal2004},
while those at 23, 111 and 600 MHz are from the recent work of
\citet{popovetal2006}. From our
non-detection at 100 MHz, we derive an upper limit of $\sim$5
kJy for the bright giant pulses
detectable at this frequency, by assuming a simple $\nu^{-2.6}$
scaling of our measured \Tsky
and nominal values for the anticipated system sensitivity. This
is much larger than the 0.7 kJy value for the brightest observed 
pulse reported by \citet{popovetal2006}. However, this
discrepancy can be attributed to the fact
that it is based on a much shorter observing duration (15 min)
and hence may not have been quite sensitive to the tail end of 
the giant pulse power-law distribution. It is still interesting that
the measured peak amplitudes of the brightest giant pulses in one hour
follow roughly power-law trends with
frequency, with an empirical relation of the form $\nu^{-4.3\pm0.2}$
at frequencies $\ga$ 300 MHz (i.e., comparable to the nominal spectral 
slope of giant pulse energies), and $\nu^{+2.7\pm0.1}$ at lower 
frequencies (if we exclude the 111 MHz value). While it is harder to
extract any meaningful information about the nature of the
spectrum from these measurements alone,
this plot may still serve as a useful guide to assess the
detection prospects of bright giant pulses. Characterization of the 
spectrum requires reliable measurements
of average giant pulse fluxes or
energies, which are not easily obtainable for all these
frequencies. Clearly, scattering will
strongly influence the giant pulse detection prospects within the
MWA-LFD's frequency range. We discuss
the detection prospects for the full system in \S~\ref{s:full}.

\subsection{Implications of Measured Pulse Broadening}

Our pulse deconvolution yields a \taud of $670 \pm 100$ $\mu$s
for the PBF \gone, and $395 \pm 50$ $\mu$s for the PBF \gtwo. As described
in \citet{bhatetal2004}, in principle, the figures of merit of the 
CLEAN deconvolution technique can be used to assess which PBF better
fits the data. Specifically, their parameter, $f_r$, which is a measure of
positivity, can be 
treated as a useful indicator of ``goodness'' of the CLEAN subtraction.
We expect $f_r ~ \la ~ 1$ for successfully deconvolved pulses, while
larger values imply slightly overCLEANed pulses. For the results in 
Fig.~\ref{fig:clean}, we obtain values of 1.3 and 0.85 respectively
for deconvolution with \PBFone and \PBFtwo, which would mean \PBFtwo is 
probably a better fit to the data. 
As \PBFtwo represents a scattering geometry with distributed material, 
this suggests that much of the observed scattering is possibly due to
the intervening ionized gas along the Earth-pulsar line-of-sight.

Our \taud estimates for the two PBFs differ roughly by a factor two, and 
this  is because \taud has different meanings for the two scenarios.
For \PBFone, \taud is
both $e^{-1}$ point of the distribution and the expectation
value of $t$, whereas
for \PBFtwo, \taud is close to the maximum of the distribution, which is 
at $(\pi^2/10)\taud=0.99~\taud$, while the expectation
value of $t$ is $(\pi^2/2)\taud=4.93~\taud$ (see eqn.~\ref{eq:pbfs}).

The measured pulse broadening can be used to infer the
scattering measure (SM), defined as $\SM \equiv \int _0 ^D ds\,
\cn (s) $, where \cn ~is the spectral coefficient of the
wavenumber spectrum of electron density irregularities in the
ISM and D is the Earth-pulsar distance. It signifies the total
amount of scattering along the line-of-sight and depends
critically on the nature of distribution of the scattering 
material. It can be related to pulse broadening via the relation
$ \taud \approx 1.1\, \Wtau~\SM^{6/5}\nu^{-4.4}D $,
where $\nu$ is in GHz, D is in kpc, and \Wtau~is a geometric
factor that depends on the line-of-sight distribution of 
scattering material \citep{cordes-rickett1998}. Using this 
relation, and assuming $\Wtau$ = 1 , we estimate the {\it effective}
SM for a uniform medium as $ (1 \pm 0.3) \times 10^{-3}$ \smu.
This can be compared to a value $1.7 \times 10^{-3}$ \smu as
predicted by the NE2001 model \citep{ne2001}.

\begin{figure}[]
\epsscale{1.0}
\plotone{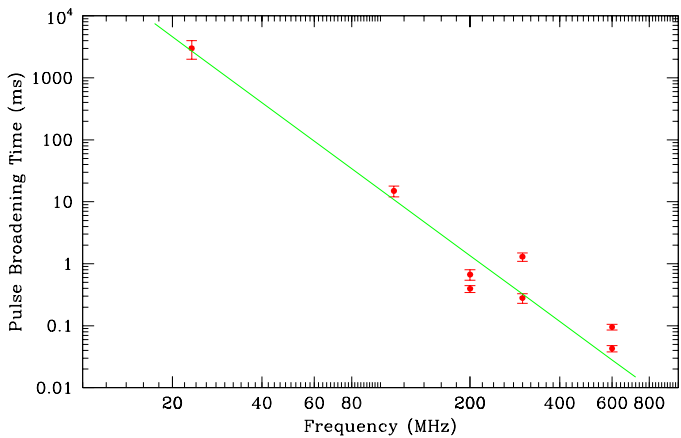}
\caption{Low frequency measurements of pulse broadening time from recent observations
of the Crab pulsar. Observations at 23, 111 and 600 MHz are from \citet{popovetal2006},
while those at 300 MHz and the higher value at 600 MHz are from \citet{sallmenetal1999}.
The solid line shows our best-fit scaling ($ \taud \propto \nu^{-3.5\pm0.2} $) to the 
data (excluding higher values at 300 and 600 MHz), which is significantly shallower than 
the $\nu^{-4.4}$ scaling expected from a Kolmogorov model. Our measurements (at 200 MHz) 
are $\sim$3--5 times smaller than that predicted based on the published measurements 
and frequency scaling.}
\label{fig:scaling}
\end{figure}

While the estimates of pulse broadening time are, in general,
reliable measures of the integrated scattering due to the ISM, 
some caution is needed while interpreting the measurements for 
the Crab pulsar.
This is because the material within or around the nebula can 
potentially influence, or may even dominate, the observed 
scattering on some (or most) occasions. 
In fact, observations have shown that the dispersion 
and scattering measurements of the Crab vary significantly with
time \citep{rankin-counselman1973,lyne-thorne75,isaac-rankin77,backer2000,lyneetal2001}.
The extreme cases are the so-called ``scattering events"
where the pulse-broadening time increased dramatically (by two
orders of magnitude) and subsequently decreased to normal over
a period of several months \citep{lyne-thorne75}. 
Such anomalies were interpreted in terms of an increase in
turbulence or density within the nebular region. 
Barring such exceptional cases, the measurements of
pulse-broadening time at any given frequency still show large variations 
from one epoch to the other. For example, \citet{sallmenetal1999} reported
the broadening time changing from 0.28 to 1.3 ms from their Green Bank 
observations at 300 MHz. More recently, \citet{popovetal2006} 
measured a value of $ 43 \pm 5 $ $\mu$s at 600 MHz, which is a factor two
smaller than that reported by \citet{sallmenetal1999} at this frequency. 
Fig.~\ref{fig:scaling} shows a summary of these measurements along with 
those from our observations. Incidentally, our measurements are $\sim$3--5 
times smaller than that predicted based on the published measurements
and frequency scaling. This may suggest that the nebular contribution 
to the total scattering is now smaller compared to those in earlier 
observations. In other words, the bulk of our observed scattering is 
likely resulting from the distributed ISM along the line-of-sight.
The best-fit frequency scaling of $\taud \propto \nu^{-3.5\pm0.2}$ 
derived by us (Fig.~\ref{fig:scaling}) is in agreement with that 
of \citet{popovetal2006} but is a significant departure from a 
Kolmogorov scaling ($\nu^{-4.4}$) favored by some of the earlier 
observations \citep{isaac-rankin77,sallmenetal1999}, when the bulk
of the observed scattering was attributed to material either inside 
or around the synchrotron nebula. 

In summary, our measurement of a low scattering and an SM estimate
that is somewhat lower than that predicted by the NE2001 electron 
density model (which largely accounts for the scattering due to the 
distributed ISM), suggest that the contribution from the nebula 
itself is probably at its lowest. This is further supported by a 
reasonably good fit obtained with \PBFtwo which corresponds to a 
distributed scattering geometry. If the scattering due to the nebula 
was dominant, one would have expected the measured pulse, and the
implied PBF, to have quite different shape given the filamentary 
structures of the nebular material. Such filaments would mean 
scattering structures with  large axial ratios, which may produce 
PBFs that are of the form ${\rm \propto t^{-1/2} exp(-t)}$ and 
possibly truncated beyond certain times \citep[e.g.][]{cordes-lazio2001}. 
Interestingly, this measurement of low scattering is also consistent 
with our recent observations at a frequency of 1400 MHz made with the 
ATCA, where we estimate the pulse broadening to be $\la$ 1 $\mu$s 
\citep{bhatetal2007}. Finally, we believe such a low scattering is 
what enabled a successful detection of the giant pulses with our ED 
system. For instance, if the scattering was to be dominated by the 
nebula, increasing the net pulse broadening by a factor 5 to 10, then 
the peak amplitude of the pulse will be reduced by a similar factor, 
rendering a positive detection rather difficult with our simple 
equipment. All in all, the measurement of a low scattering and our 
detection prospects are coupled, and perhaps it is a coincidence that 
our observations were made at an epoch when the scattering 
happened to be unusually low.

\section{Giant Pulse Detection Prospects with the Full MWA-LFD System} \label{s:full}

Our successful detection of the Crab giant pulses with the ED system opens up some exciting
prospects for the studies of giant pulses and other fast radio transients with the full MWA-LFD
system when it becomes operational. The full system will have a sensitivity that is
over two orders of magnitude larger than the ED system, and in combination with a much larger
bandwidth feasible for data recording and the instrument's unique capability to observe at more
than one frequency simultaneously, the MWA-LFD will offer some brand new avenues of investigation 
in this arena.

\begin{figure*}[t]
\epsscale{2.0}
\plotone{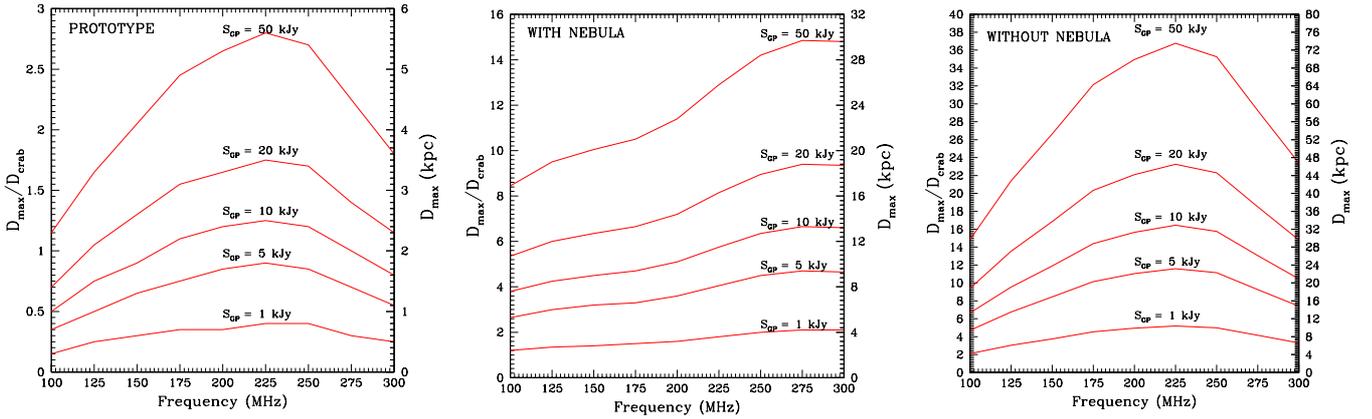}
\caption{Giant pulse detection sensitivity of the MWA-LFD system. The sensitivity is
expressed as maximum distance Dmax to which a giant pulse is potentially detectable.
For the prototype system ({\it left panel}), the achievable sensitivity is dominated 
by the system noise itself, and an optimal performance is expected near 200 MHz. The 
middle and right panels show projected sensitivities for the full system. The nebular 
emission is important for ``Crab-like'' pulsars that reside in ``Crab-nebula-like'' 
environment ({\it middle panel}). The nebular contribution to the system noise decreases 
significantly above 190 MHz as the nebula is resolved by the array beam (assuming an 
array diameter  of $\approx$1 km), thus significantly increasing the system sensitivity 
at these frequencies. For field pulsars however ({\it right panel}), Dmax is largely 
determined by the sensitivity of the array itself, with the best performance expected
near 200 MHz.}
\label{fig:sens}
\end{figure*}

Fig.~\ref{fig:sens} shows giant pulse detection prospects of the MWA-LFD system, where 
the sensitivity curves for the full system are shown (middle and right panels) along 
with those for the prototype ED system (left panel). 
The sensitivity curves are dramatically different for the two cases, which is due to an 
interplay of different factors involved in the calculations. 
For the ED system, the system noise dominates over all other contributions and a maximum 
in sensitivity is expected near 200 MHz.
For ``Crab-like'' pulsars that reside in ``Crab-nebula-like'' environment, the sensitivity 
calculations take into consideration factors such as, contribution from the nebular 
emission to the system noise, as well as frequency dependencies of the basic system 
performance parameters (i.e., \Tsys and G) and the sky background (\Tsky). 
The combination of a decreasing  \Tsky and a reduced nebular contribution as the nebula 
becomes resolved at higher frequencies make up for the decrease in tile gain and a 
larger \Tsys expected at higher frequencies. 
This effectively yields higher sensitivity at frequencies $\ga$200 MHz. 
Naturally, much higher sensitivities can be achieved for field pulsars with no nebular 
background, where \Dmax is largely determined by the system performance parameters such 
the array gain and noise temperature. 

To elaborate further on our sensitivity calculations, the
radiometer equation can be re-written
for the case of giant pulse detection as (see e.g.
\citet{maura-cordes2003,cordesetal2004}),
\be
{\rm \left( { S \over N } \right)_{det} } = ~ {\rm D}^{-2} \left( { \Sgp \over \Ssys } \right) \left( \Npol \delnu \Wp
\right)^{{1 \over 2}} 
\label{eq:sens}
\ee
where \SNdet is the signal-to-noise achievable for a giant pulse of peak
amplitude \Sgp using a system of
sensitivity \Ssys, with \delnu being the bandwidth over which
the data are recorded and \Wp is the
effective pulse width, and \Npol is the number of polarizations
that are summed. Conversely, for
a given system with sensitivity \Ssys, the maximum distance to
which a giant pulse of amplitude \Sgp is
potentially detectable is given by
\be
\Dmax = \left( {\rm {S\over N} } \right)^{-{1\over2}} \left( { \Sgp \over \Ssys } \right)^{{1\over2}} \left( \Npol \delnu \Wp \right)^{{1\over4}}
\label{eq:dmax}
\ee

For the MWA-LFD system, \Ssys is a strong function of the
observing frequency, as both the system temperature and 
gain vary significantly with frequency
\citep{bowmanetal2006}. Further, the full array will 
comprise 512 tiles spread over a region of $\sim$1 km.
Thus, both the effective gain and resolution of the array 
(i.e., half-power width of the phased array beam pattern)
are such that \Ssys is strongly influenced by the nebula.
For an array layout extending out to 1 km, the nebula remains
unresolved at the lower half of the MWA-LFD's operating
frequency range, but becomes resolved near 200 MHz, where the
effective resolution (beam width) becomes comparable to the
Crab's characteristic diameter ($5.^{\prime}$5).
The net system sensitivity (in units of Jy) is thus given by 
$ \Ssys (\nu) = \Ssyso (\nu) + \fnu
\Scn (\nu) $, where $ \Ssyso $ is the
system sensitivity without the nebular contribution to the
system noise and \fnu is the beam dilution factor, given by 
$ \fnu = \Omega _A (\nu) / \Omega _{CN} $ where \OmgA and 
\OmgCN are the solid angle areas of the phased array beam and
the nebula, respectively.
The system sensitivity when pointed away from the nebula is 
$ {\rm \Ssyso (\nu) = \Tsyso (\nu) / G (\nu) } $,
where $ \Tsys (\nu) = \Tsky (\nu) + \Trecv (\nu) $ is the net
system temperature. Furthermore, the nebular emission is also 
frequency dependent (see \S~\ref{s:calib}), and the integrated 
flux density is expected to change as much as by 30\% over the 
MWA-LFD's frequency range.

Thus, in the lower half of the 
MWA-LFD frequency range (100--200 MHz), sensitivity is largely
determined by the combination 
of a decrease in \Tsky and an increase in the tile gain, and
then sharply increases near 
mid-way as the nebula becomes resolved (Fig.~\ref{fig:sens}). A
giant pulse as bright as the 
brightest one
in our data is thus potentially detectable with the full system
out to a distance of $\sim$17
kpc at 100 MHz and out to $\sim$30 kpc at 300 MHz. Indeed these
limits will be even larger in
the case of giant-pulse emitters that do {\it not} reside in a nebula. 
In short, bright giant pulses are potentially detectable over a wide 
range of Galactic distances within the full range of MWA-LFD frequencies.

The brightest giant pulse detected in our data (\Sgp = 50 kJy) has an 
energy content of $\sim1.85 \times 10^5$ kJy $\mu$s. 
Although our ED system is not capable of detecting the integrated emission
from the pulsar, assuming a pulsar flux density of $\sim650$ mJy at 400 MHz 
and a spectral index of $-$3.1 \citep[e.g.][]{manchesteretal2005}, it turns 
out such a giant pulse will be at least $\sim100$ times more energetic than 
the average emission (if we assume the spectrum does not turn over until 
$\sim$200 MHz). 
Given this, and an expected two orders of magnitude improvement in sensitivity 
for the full array, it is quite likely that almost all pulses emitted through
the giant-pulse emission mechanism (say, energy $\ga 10 \times$ the average
emission) will be potentially detectable with the full system at 200 MHz. This 
will enable a detailed characterisation of the giant pulses and their energy 
distributions at low frequencies, a poorly 
studied aspect of the Crab giant pulses.

It is worthwhile to examine the detection prospects for other known
giant-pulse--emitters such as PSRs B0540--69 and B1937+21. PSR B0540--69
is a young, Crab-like pulsar located inside the Large Magellanic Cloud 
(D $\sim$ 50 kpc), for
which the giant-pulse energy distribution is known to follow the form
$ P ( E > E_o ) = K E_o ^{-\alpha} $; where P is the probability of emitting
a giant pulse with energy greater than $E_o$ (in units of the average pulse
energy), and $K$=0.26 and $\alpha \sim 1.8$
from \citet{johnstonetal2004}. The pulsar period is 50.35 ms and thus the
strongest pulse emitted in 1 hr will have an energy $E_o \sim 235 \langle E \rangle$
where $\langle E \rangle$ is the average emission. Assuming giant pulses 
dominate the pulsar emission and a spectral index of $-$3.6 (and no spectral turnover 
down to $\sim$200 MHz), we estimate the pulsar flux to be 27 mJy at 200 MHz, thus a 
giant pulse with energy 235$\langle E \rangle$ would imply $3.3 \times 10^2$ kJy $\mu$s. 
This is 56 times weaker than our brightest giant pulse, or 20 times below the detection 
threshold of the ED system. However, with the full array ($\sim$170 times improvement 
in sensitivity), such a giant pulse should be detectable with S/N $\sim$ 85. And given 
the $K E_o ^{-\alpha}$ form for the energy distribution, we estimate $\sim 50$ pulses 
to be potentially detectable in 1 hr of observation. 

For the millisecond pulsar B1937+21, the giant-pulse energies follow a similar 
power-law form with K=0.032 and $\alpha \sim 1.8$ at 430 MHz \citep{cognardetal1996}. 
Following similar arguments as above, we estimate the brightest giant pulse from 
this pulsar in 1 hr will be $\sim500$ times as energetic as its average emission. 
While this is 8 times below the detection threshold of our ED system, it will be 
easily detectable (S/N $\sim$ 212) with the full array and assuming a power-law index
of 1.8, we estimate there to be as many as $\sim$ 250 giant pulses with S/N $>10$ in 1 hr 
duration. While these are only some rough estimates, and do not take into consideration
the impact of increased scattering at low frequencies, it is evident that the full system 
offers exciting prospects for giant pulse studies in general.

While the above calculations are specifically performed for
giant pulses, 
the same treatment also applies to fast radio transients of similar
characteristics (i.e., time scales of the order of 100 $\mu$s or longer). The
superb RFI-quiet environment of the proposed MWA-LFD site can be exploited to 
take the best advantage of this potential. Moreover, 
the instrument's capabilities such as multibeaming will serve
as powerful discriminators of real versus spurious signals. 
Thus the full system is indeed a promising instrument for science related to 
giant pulses and similar fast radio
transients at low frequencies. 

\section{Conclusions and Future Work}

Using a simple equipment operating in the Western Australian
outback, we have successfully detected
giant pulses from the Crab Nebula pulsar at a frequency of 200
MHz. Despite a large sky background severely limiting the
sensitivity achievable at such low frequencies, our system, 
comprising just three tiles, each consisting of a
4 x 4 array of dipoles, detected a total of 31 giant pulses 
over a duration of 3.5 hours. The measured pulse shape is 
significantly broadened by
multipath scattering due to the ISM, and results
in a degradation of the intrinsic pulse amplitude by nearly a
factor of two.
The deconvolution of the measured pulse using a CLEAN-based
procedure yields a pulse-broadening time of 670$\pm$100 $\mu$s 
for the case of a thin slab scattering
model, and 400$\pm$50 $\mu$s for a model where the scattering 
material is uniformly distributed along the line of
sight. The implied degree of scattering is the lowest that has been
reported towards the Crab pulsar from
observations made so far.
In fact, our detection of giant pulses in such large numbers
would not have been possible but for such low level of 
scattering. Together with recent observations at low 
frequencies, our measurements of the pulse-broadening time 
support a shallow 
scaling in frequency ($ \taud \propto \nu^{-3.5\pm0.2} $).
With the sensitivity of our equipment (a gain of $\sim$0.01
\kpjy and a system temperature of $\sim$400 K towards the Crab),
pulses that are brighter than $\sim$9 kJy in amplitude are
easily detected.
The brightest giant pulse detected in our data has a peak
amplitude of $\sim$50 kJy and a width of $\approx$300 $\mu$s,
and the equivalent brightness temperature is $10^{31.6}$ K,
assuming a pulsar distance of 2 kpc.

The success of our exploratory observations underscores the
tremendous potential the full MWA-LFD system will offer for 
studies of pulsar science in general, and
for giant pulses and fast transients in particular, at low 
radio frequencies. With a sensitivity that is over two orders 
of magnitude larger
than that of the prototype equipment, the full system will have 
the capability to detect giant pulses over a wide range of 
Galactic distances.
For instance, giant pulses as bright as $\sim$50 kJy will be 
potentially detectable out to a distance of  $\sim$30 kpc in
the case of Crab-like pulsars, and even further in the case
of objects that do not reside in a nebula.
In addition to enabling an in-depth study of Crab giant pulses at
frequencies $\la$300 MHz, the MWA-LFD will be a promising 
instrument for a wide variety of pulsar and transient science at 
low frequencies.

\bigskip

\noindent
{\it Acknowledgements:} 
Data processing for the work presented here was carried out at the Swinburne supercomputing
facility, and data storage resources were provided by the GrangeNet facility, Canberra, ACT. 
We thank S. Tingay and Y. Gupta for valuable discussions pertaining to the phased-array 
realisation of the ED array, and C. West and P. Fuggle for assistance with the GrangeNet 
access by the Swinburne supercomputer. 
The prototype field deployment effort for the MWA-LFD project was supported by the MIT School 
of Science, the MIT Haystack Observatory, University of Melbourne, Australian National 
University, Curtin University, Australian National Telescope Facility, University of Western 
Australia, Harvard-Smithsonian Center for Astrophysics, Mileura Cattle Company, the government 
of Western Australia, the Australian Research Council and the U.S. National Science Foundation.


\end{document}